\documentclass[prd,aps,twocolumn,preprintnumbers,showpacs,nofootinbib,superscriptaddress,notitlepage]{revtex4-1}
\usepackage{hyperref}

\usepackage{bm}
\usepackage{amsmath}
\usepackage{slashed}
\usepackage{xcolor}
\usepackage{graphicx}
\usepackage{caption}
\usepackage{subcaption}
\usepackage[normalem]{ulem}
\usepackage[utf8]{inputenc}
\usepackage[T1]{fontenc} 

\newcommand{\beq}{\begin{eqnarray}}
\newcommand{\eeq}{\end{eqnarray}}

\begin{document}

\title{Euclidean Effective Theory for Partons in the Spirit of Steven Weinberg}

\author{Xiangdong Ji}
\email{xji@umd.edu}
\affiliation{Maryland Center for Fundamental Physics,
Department of Physics, University of Maryland,
College Park, Maryland 20742, USA}

\date{\today{}}
\begin{abstract}

The standard formulation of parton physics involves light-cone correlations of quark and gluon fields in a hadron, which leads to a widespread impression that it can only be studied through real-time Hamiltonian dynamics or light-front quantization, which are 
challenged by non-perturbative computations with a pertinent regulator for light-cone/rapidity divergences (or zero modes). As such, standard lattice QCD studies have been limited to indirect 
parton observables such as first few moments and short-distance correlations, which do not provide the $x$-distributions without solving the model-dependent inverse problem. 
Here I describe an alternative formulation of partons in terms of equal-time (or Euclidean) correlators, which allows to compute precision-controlled $x$-distribution through lattice QCD simulations. This approach is in accord with Weinberg's pioneering idea of effective field theory as well as Wilson's renormalization group, in which the large hadron momentum serves as a natural cut-off for light-cone/rapidity divergences and can ultimately be eliminated through a method like the ``perfect action'' program in lattice QCD.  

\end{abstract}
\maketitle

\section{Weinberg and the Rise of Effective Field Theories}
Having established the electroweak unification theory of elementary particles based on the renormalizability constraint~\cite{Weinberg:1967tq}, Steven Weinberg reversed his attitude toward renormalization, and argued that non-renormalizable terms in a lagrangian can still have field-theoretical significance so long as their divergences can be controlled through a systematic expansion of some small parameters (power counting). If the short-distance/large-momentum contributions can be lumped into renormalized constants of finite numbers in a given order (of power counting), the long-distance quantum fluctuations can still be calculated and tested through combinations of experimental observables, independent of these constants. The case in point is chiral effective field theory (EFT) of strong interactions in which low-energy meson and hadron structure and scattering physics can be calculated in terms of a fixed number of ``low-energy'' constants encoding ultraviolet (UV) physics at a fixed order, which can in turn be determined by fitting to experimental data~\cite{Weinberg:1968de,Weinberg:1978kz,
Gasser:1983yg}. A similar approach has been advocated by Weinberg to study nuclear physics, known as nuclear EFT, replacing nuclear models which have been the main-stream approach since the beginning of the field
~\cite{Weinberg:1990rz,Weinberg:1991um,
Kaplan:1996xu,Epelbaum:2008ga}.  

Wilson's renormalization group and operator expansion developed  for high-energy scattering is a similar factorization of physics at different scales into local operators and coefficient functions, in which the high-energy physics encoded in the Wilson coefficients can actually be calculated in quantum chromodyanamics (QCD) due to asymptotic freedom, whereas the infraed (IR) physics in operator matrix elements can be calculated non-perturbatively through low-energy QCD~\cite{Wilson:1969zs,Christ:1972ms}. This can be formulated as an application of EFT, but the exact meaning of which depends on one's focus: For high-energy physicists, perturbative QCD is an EFT with low-energy physics parametrized as ``high-energy constants'' in the sense of Weinberg. For the hadron structure theorists, non-perturbative QCD is the EFT after integrating out the high-energy degrees of freedom in hard-scattering cross sections.

Weinberg had been perhaps my greatest physics hero throughout my scientific career. He was a rationalist, giving others the illusion that if we work ten times as hard as we do, we can become another Weinberg. I first heard about him when I was a graduate student at Peking University in 1982 when one of my classmates was reading his famous book on ``Gravitation and Cosmology.'' My first real life encounter at a distance was when I was a junior faculty at MIT at a workshop on chiral dynamics in June 1994, in which he gave a talk about his EFT approach to nuclear physics. It was striking to me at that time that a renowned theorist like him was still interested in century-old problems of nuclear forces and pion scattering etc, which seemed to have gone out of fashion even in nuclear physics. The most memorable face-to-face conversation was when he came to U. Maryland to give a nuclear seminar around the turn of the century in which he talked about the EFT approach to Nambu-Jona-Lasino type of models. Again, his presentation was so logical and persuasive that if someone just follows the nose, he/she would have obtained the same results, a sense of ``inevitableness'' in his frequently-used terms. Weinberg was at one time the most-cited physicist in the world. Yet it was interesting that during his visit at UMD, he complained that his work on ``mended symmetry'' seemed to have been ignored by others~\cite{Weinberg:1990xn}. He described it as ''a child that has not grown so well." Over the years, I had a number of email communications with him for various reasons, and what impressed me was that he always replied to emails promptly and never left them unanswered, which is quite rare among Nobel Laureates.

Half-century passed, EFT has become a most basic and yet a most essential tool of theoretical physics (see, for example, ~\cite{Georgi:1993mps,Manohar:2018aog}).
Many EFTs have been estalished, but the spirit is same: scale separation, power counting, and 
inevitableness of results as consequences of general principles given the relevant degrees of 
freedom and symmetries. While some EFTs are invented to make calculations easier, like non-relativistic quantum electrodynamics (NRQED), others are essential to establish the existence of very theories and their calculability.  In the latter case, we mention the renormalization program itself, i.e., finding proper UV regulators of quantum field theories, without which the latter are not well-defined. 
In electro-weak theory (also in perturbative QCD), dimensional regularization has played the key role for performing high-precision perturbative calculations. For low-energy QCD, on the other hand, lattice regularization is essential to define the theory non-perturbatively, which is an EFT of the continuum theory through perfect actions~\cite{Symanzik:1983dc,Symanzik:1983gh}. Similarly, finding a non-perturbative light-cone regulator is essential to solving the EFT for
parton dynamics.

\section{QCD Factorization and Parton as EFT Objects}

Asymptotic freedom of QCD allows calculating physics at large momentum scales in standard Feynman perturbation theory. However, no physical observables involve only perturbative physics: non-perturbative physics at scale $\Lambda_{\rm QCD}$ is always present in any physical processes. Therefore, in general, one can perform scale separation so that high-momentum physics is treated perturbatively, whereas the IR physics is parametrized by non-perturbative matrix elements. Schematically, one can decompose, for example, a quark field into
\begin{equation}
    \psi(x) = \psi_{\rm high} (x) + \psi_{\rm low}(x) \ , 
\end{equation}
and the low-energy dynamics is represented by matrix elements of $\psi_{\rm low}$'s. The application of perturbative QCD (pQCD) can be viewed as an example of Weinberg's EFT with high/low energy calculability switched. 
Compared with chiral perturbative theory, however, pQCD has two major differences: 1) Due to light-cone dominance in high-energy processes, power counting is not the same as dimensional counting. Instead one has now the so-called twist counting (see for example ~\cite{Peskin:1995ev}). The result is an infinite number of ``high-energy constants'' which are functions of hadron momentum fraction $x$ and parameters, forming sundry light-cone/parton distribution functions (PDFs). 2) Apart from determining these constants from (global analyses of) experimental data (for example ~\cite{NNPDF:2017mvq,Hou:2019efy}), they are in principle calculable directly in low-energy QCD. 

To phenomenologically fix an infinity number of high-energy constants or a parton distribution function (PDF) needs in principle an infinite number of data. However, we do not have infinite or even systematically discretized data sets covering all phase space, and therefore it is often labelled as ``inverse problem'' in the literature, meaning no unique solution is possible. One can find a set of probable solutions so that known data are well-described. A general strategy is to assume specific correlations between the infinite number of constants or that PDFs have some guessed forms of $x$ dependence~\cite{Hou:2019efy}. Similar approaches have been used to fit the first few moments and/or short-distance correlations from lattice QCD with model functions. However, these approaches do not allow fixing PDFs at any particular $x$ with rigorously-quantifiable systematic error.

Therefore, it is imperative to find a way to calculate at least some of ``high-energy constants'' with controlled systematics, which in turn requires a better understanding of the physics content of partons. Using the EFT approach, one can integrate out $\psi_{\rm high}$ first, and the remaining low-energy theory involves then $\psi_{\rm low}(x)$ only. As has been known since the 1970's, PDFs are Fourier transformations of the gauge-invariant matrix elements of the light-cone correlations of $\psi_{\rm low}(x)$, and these degrees of freedom are Feynman's partons. 

There are two apparently-different formalisms for parton physics that are equivalent. In the first one, a hadron can have any three-momentum (in particular in the rest frame $\vec{P}=0$), and collinear fields with coordinates along the light-cone directions are probed. There are also soft degrees of freedom which cannot be distinguished 
from the collinear fields at $x=0$. In perturbative calculations, the light-cone correlators will generate additional divergences beyond the usual ones. An effective lagrangian can be explicitly
introduced to describe the dynamics of the soft and collinear fields. The resulting EFT is called soft-collinear effective theory (or SCET). For example, the leading effective lagrangian for the collinear quark fields~\cite{Bauer:2000ew,Bauer:2000yr} is, 
\begin{eqnarray}
{\cal L}_{\rm SCET} &&= \overline {\psi}_{k}(x) \left[i\bar{n}\cdot D \right. \nonumber 
\\ && \left. + (i\slashed{D}_\perp) \frac{1}{{in\cdot D}}  (i\slashed{ D}_\perp ) \right] \frac{\slashed{n}}{2}\psi_{k'}(x) + ... \ , 
\end{eqnarray}
where ${\bar n}=(1,0,0,1)$ and $n=(1,0,0,-1)$ are conjugate light-cone vectors. The hadron is moving along positive $z$-direction, and $k^\mu=x{\bar n}^\mu + k_\perp^\mu $ and $k'^\mu=x' \bar{n}^\mu + k_\perp '^\mu $ are momentum labels of collinear quark fields (we have not factored out the rapid oscillating phases here). Similar terms can be added for the collinear gluon fields, as well as soft fields. 
Note that due to the inverse differential operator $1/in\cdot D$  (corresponding to a Wilson line), the lagrangian density is non-local and simplifies when choosing $n \cdot A=0$ gauge.
The collinear PDF's are the matrix elements of the light-cone correlators of the collinear fields. When calculating them in covariant perturbation theory, dimensional regularization is sufficient for divergences from light-like separations. An explicit light-cone regulator has to be introduced for transverse-momentum dependent quantities. There are a number of proposals for light-cone regulators for perturbative calculations in the literature (see a summary in ~\cite{Ebert:2019okf}).  When working to all orders in perturbation theory, 
a perturbative regulator is sufficient for defining a scheme for non-perturative matrix elements.  
Calculating non-perturbative parton physics with SCET has not been attempted so far, but explicit time-dependence is clearly a key obstacle.

An older formulation of parton EFT is based on the so-called light-front quantization~\cite{Dirac:1949cp,Brodsky:1997de}, in which the Hamiltonian $H_\infty$ and Fock space can also be derived in the infinite momentum limit of time-independent (old-fashioned) perturbation theory~\cite{Weinberg:1966jm}.
In this case, both observables and hadron states are constructed out of collinear particles with one infinite-momentum direction, referred to as parton degrees of freedom. All others with finite momentum can be lumped into zero modes. To keep light-cone non-locality simple, it is standard to choose the light-cone gauge $n\cdot A=0$. However, $H_\infty$ is still a non-local operator of light-cone distributed fields, which has various light-cone divergences. Compared with SCET, there are extra light-cone divergences from the light-cone gauge as well as splitting of covariant Feynman diagrams into light-cone-time-ordered ones. Renormalizing these divergences properly for non-perturbative calculations has been a great challenge, because it is very difficult to restore all symmetries that are present before the infinite momentum limit is taken. It is fair to say that no satisfying light-cone regulator has been found for light-front QCD, which hinders any attempt to calculate PDF's from first principles. 

Light-front quantization can formally be obtained from integrating out the energy or time in the ordinary or SCET Feynman diagrams. 

\section{Factorization of DIS in terms
of Euclidean Correlators}

To develop a practical method for computation of parton physics, 
we will demonstrate, using the example of well-known 
deep-inelastic scattering (DIS), that high-energy processes 
can be factorized in terms of {\it Euclidean correlators} of collinear fields, which means that partons can be formulated in Euclidean field theory. 

To do so, we use the so-called Bjorken (or Breit) frame in which 
the virtual photon momentum has only a non-vanishing $z$ component and the hadron is moving with velocity $v^\mu$ in the same direction. The hadron and virtual photon momenta are
\begin{eqnarray*}
   q^\mu &=& (0,0,0,-Q) \ , \nonumber \\
   P^\mu &=& M\gamma v^\mu \ ,~~~~ \gamma = \sqrt{1+\frac{Q^2}{4x_BM^2}}  \ ,   
\end{eqnarray*}
respectively, where $v^\mu =  (1, 0,0,v)$ and $v^\mu v_\mu =1/\gamma^2$. In the Bjorken limit, $0<x_B<1$, $\gamma\sim Q\to \infty$ and $v\to 1$.

Let us first consider the hand-bag diagram shown in Fig. 1, in which 
the hadron tensor is,  
\begin{eqnarray}
    W^{\mu\nu}(x_B, Q^2)  && = \frac{1}{2\pi}{\rm Im} \int  i \frac{d^4k}{(2\pi)^4}
                  {\rm Tr}\left[\gamma^\mu S(k+q) \gamma^\nu M(k)\right]  \nonumber \\ && + ~~{\rm crossing}
\end{eqnarray}
where $S(k)$ is the single quark propagator of four-momentum $k^\mu$, and $M(k)$ is the single quark 
Green's function in the hadron, 
\begin{equation}
          M(k)^{\alpha\beta} = \int d^4\xi e^{i\xi\cdot k}\langle P|T\overline{\psi}_{\rm low}^\beta(0)\psi_{\rm low}^\alpha (\xi)|P\rangle
\end{equation}
where $|P\rangle$ is the hadron state.

\begin{figure}[!th]
\centering
\includegraphics[scale=0.4]{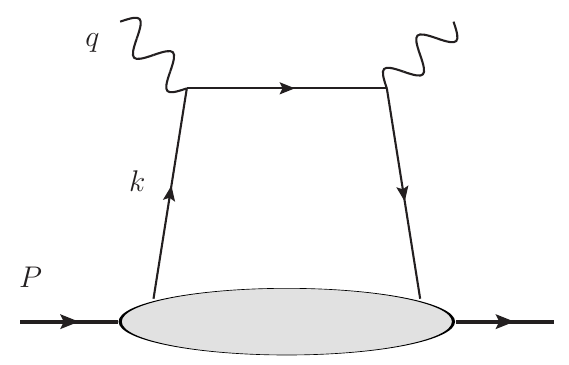}
\includegraphics[scale=0.4]{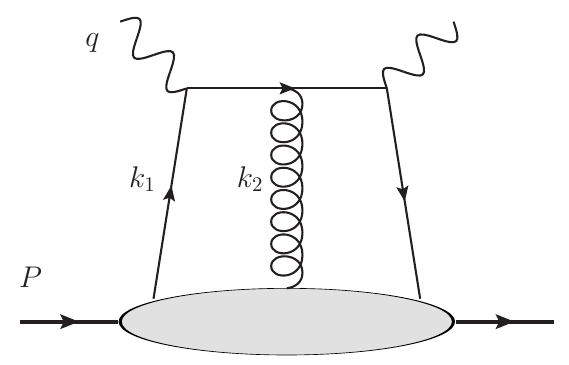}
\caption{Tree-level diagrams for DIS process. }
\label{fig:Fig_quasiTMDPDF}
\end{figure}

We will now restrict $\psi_{\rm low}$ to those collinear fields making up the hadron with velocity $v$, 
\begin{equation}
\psi_{\rm low}(x) = \psi_v (x) + ... \ , 
\end{equation}
then
\begin{equation}
          M(k)^{\alpha\beta} =  \int d^4\xi e^{i\xi\cdot k}\langle P|T\overline{\psi}_v^\beta(0) \psi_v^\alpha(\xi)|P\rangle \ . 
\end{equation}
The effective Fourier components of $\psi_v (x)$ have momentum $k^\mu$, with the following decomposition, 
\begin{equation}
   k^\mu = \alpha v^\mu + \beta \bar v^\mu + k_\perp^\mu  ,~~~ k^2 \sim \Lambda_{\rm QCD}^2, 
\end{equation}
where $\bar v^\mu = (v, -1)$, $\bar v_\mu^2=-1/\gamma^2$, and $v\cdot \bar v = 2v$; $\alpha\sim \gamma \Lambda_{\rm QCD}$ 
and $\beta\sim \Lambda_{\rm QCD}/\gamma $. Thus the coefficient of $\bar v^\mu $ are suppressed by $1/\gamma$. 
Moreover,  $\psi_v$ satifies, 
\begin{equation}
         \slashed{v} \psi_v = 0  \ , 
\end{equation}
following from the leading order equations of motion (EOM) in $1/\gamma$. 

The leading contribution to the hadron tensor comes from transverse polarization of the photon and thus $i, j = \perp$. In light of the trace in Eq.(3), the quark propagator can be simplified,
\begin{equation}
S(k+q) = \frac{i(\slashed{k} +\slashed{q})}{(k+q)^2 + i\epsilon} = \frac{i\slashed{q}}{2k\cdot q -Q^2+i\epsilon} = \frac{i\gamma^z}{2k^z - Q + i\epsilon} \ , 
\end{equation}
where in the second equality, we use Eq. (8) in the numerator to eliminate $\slashed{k}$ and neglected $k^2\sim \Lambda_{\rm QCD}^2$ in the denominator. 
Defining $k^z = Qy/2x_B$,  the integration over $k^0$ and $k_\perp$ in Eq. (3) can be carried out, 
\begin{eqnarray*}
    W^{\mu\nu} && = - g^{\mu\nu}_\perp {\rm Im} \int^\infty_{\infty} \frac{dy}{2\pi}
       \tilde f(y) \frac{1}{y/x_B-1+i\epsilon} + {\rm crossing} \nonumber \\
        &&= - g^{\mu\nu}\frac{1}{2}\left(\tilde f(x_B) + \tilde f(-x_B)\right) \ ,  
\end{eqnarray*}
where 
\begin{eqnarray}
    \tilde f(y, P^z) && = \frac{1}{2} \int dz  e^{izk^z}\langle P|{\bar \psi}_v(z)\gamma^z \psi_v(0)|P\rangle \nonumber \\
      && = \frac{1}{2P^z} \int d\lambda   e^{iy\lambda}\langle P|{\bar \psi}_v(z)\gamma^z \psi_v(0)|P\rangle  \ , 
\end{eqnarray}
with dimensionless $\lambda \equiv zP^z$. The above result is identical to the standard QCD factorization result, except that the 
distribution $\tilde f(y,P^z)$ replaces the light-cone distribution
\begin{equation}
   f(x)  = \frac{1}{2P\cdot n} \int^{\infty}_{-\infty} d\lambda   e^{ix\lambda}\langle P|{\bar \psi}(\lambda n)\gamma^+ W(\lambda n, 0) \psi(0)|P\rangle \ . 
\end{equation} 
The key of the derivation is that the $k^0$ components of the quark four-momentum can be eliminated through the equation of motion (EOM) of the effective field $\slashed{v}\psi_v$=0. And therefore, $k^0$ integration can be carried out 
in the hadron matrix element, resulting in an equal-time correlation function, which
is the ordinary momentum distribution of quarks, the same as in non-relativistic systems.

The above derivation can be repeated for the diagrams where the quark is propagating in the background gluon field of the hadron, which we will call $A_v^\mu$. The collinear gluon field can also be decomposed in the same manner, 
\begin{equation}
  A^\mu_v  = \alpha v^\mu + \beta \bar v^\mu + A_\perp^\mu \ , 
\end{equation}
and at leading order, only the $\alpha$ component dominates. 
For example, when there is one interaction with the gauge potential (see Fig. 1), we replace the quark propagator in Eq. (3) by 
\begin{equation}
      S(k_1+q) \gamma^\alpha S(k_1+k_2+q) \ ,  
\end{equation}
and $M(k)$ by $M^\alpha(k_1,k_2)$ where an additional $A_v^\alpha(k_2)$ appears. 
Both $S$ can be simplified by the EOM of the effective field, 
\begin{eqnarray}
   && S(k_1+q) \sim \frac{i\gamma^z}{2k_1^z-Q+i\epsilon}, \nonumber \\  && S(k_1+k_2+q) \sim \frac{i\gamma^z}{2k_1^z+2k_2^z-Q+i\epsilon}
\ , \end{eqnarray}
For $\slashed{A_v}$, the situation is a bit involved, 
since $A^z_v \sim A^0_v$ as $v\to c$, we have the leading contributon,  
\begin{equation}
   \slashed{A}_v = A^0_v\gamma^0 - A^z_v \gamma^z 
\end{equation}
However, after commutation with $\gamma^z$, 
we have 
\begin{equation}
   \gamma^z \slashed{A}_v = -(A^0_v\gamma^0+A^z_v\gamma^z) \gamma^z 
                         \sim  2A^z_v
\end{equation}
where again we have used quark's EOM and $A^z_v \sim A^0_v$. 
Therefore, effectively all $\slashed{A_v} = -2A^z_v \gamma^z$, which allows one to calculate diagrams with an arbitrary number of $A^\mu_v$ interactions. 

Adding all the quark eikonal interactions, one has 
    \begin{equation}
    \tilde f(y, P^z) = \frac{1}{2P^z} \int^\infty_{-\infty} d\lambda   e^{iy\lambda} \langle P|\bar \psi_v(z)W(z, 0)\gamma^z \psi_v(0)|P\rangle
\end{equation}
where $W(z,0)=P\exp\left(-ig\int^z_0 A^z(z')dz'\right)$ keeps the non-local Euclidean correlator gauge-invariant, and the integration is along 
the $z$-direction. The above expression is now completely gauge invariant. 

Therefore, although the final-state quark is propagating along the light-cone direction, the gauge link can be effectively chosen along the $z$-direction because the hadron and its constituent quarks and gluons
are moving with large momentum. This is true only in this class of large-momentum frames, 
not when the hadron is at rest where the light-cone correlator becomes essential.

The above discussion can easily be generalized to the Drell-Yan process in which
the virtual photon is time-like. For each event, we can choose a frame in which $q^\mu=(Q,0,0,0)$ and  
one of the hadrons is moving with velocity $v^\mu=(1,0,0,v)$ and the other with
$v'^{\mu}=(1,0,0,-v')$. The quark fields can be split into
\begin{equation}
  \psi (x) = \psi_v(x) + \psi_{v'}(x) + ...
\end{equation}
and the cross section can be written in terms of the product of two quark momentum distributions $\tilde f(x_1, P^z)$ and $\tilde f(-x_2, P'^z)$. 

Eqs. (11) and (17) differ by a Lorentz tranformation of the correlators. 
In fact, any correlator that approaches the light-cone correlation after Lorentz
boost can be used to define the non-perturbative hadron structure physics in hight-energy scattering, including time-like ones. 

\subsection{Large Momentum/Velocity/Rapidity Effective Theory}

In the above discussion, we arrived at an EFT for low-energy QCD in high-energy processes, which contains only equal time (Euclidean) correlators of collinear fields of velocity $v$. The leading effective lagrangian for the quark collinear modes can be written as, 
\begin{equation}
  {\cal L}_{q,v}^{(0)} = \overline{\psi}_v \left[iv\cdot D + 
   \frac{i {\bar v}\cdot D}{2\gamma^2}
   + (iD_\perp)\frac{1}{2i\bar v \cdot D}(iD_\perp)\right] \slashed{\bar v}\psi_v
\end{equation}
where ${\bar v} = (v,0,0,-1)/2v$ and $v_\mu {\bar v}^\mu=1$. 
One can also add the leading-order lagrangian for the gluon collinear modes. This effective theory formally converges to SCET or light-front quantization in the $v\to c$ limit. However, there are some fundamental differences which we summarized below: 

\begin{itemize}
    \item{Momentum evolution equation: The momentum distribution is frame-dependent and the large momentum dependence follows a renormalization group equation~\cite{Ji:2014gla}. But it has no light-cone divergences. In a certain sense, frame-dependence is a way to regulate the light-cone divergences which arise from interchanging the orders of the limits of $P^z \to \infty$ and $\Lambda_{UV} \to \infty$. This exchange is not legal in case of non-trivial UV properties. The limit $P^z\to \infty$ first ignoring the UV cutoff is the standard light-front/SCET approach which leads to extra divergences hard to regularize for non-perturbative calculations respecting all symmetries. In our present approach, the UV cut-off $\Lambda_{UV}$ is larger than any scale in the problem including $P^z$, and the light-cone divergences are avoided. However, due to asymptotic freedom of QCD, the two orders of limits differ only in perturbation theory but are equivalent for non-perturbative physics. The $P^z$ evolution for the second limit is equivalent to the renormalization group (RG) evolution of light-cone regularization for the first limit.   
    In the case of collinear factorization, the resulting momentum evolutions are similar to the standard RG equations for PDFs. When the transverse momentum is involved, the momentum evolution is related to the rapidity evolution in light-cone theory. Therefore, one can directly measure the momentum distributions at a particular $P^z$ through a global analysis similar to that 
    for PDFs.}
    
    \item{Non-perturbative light-cone regulator: The new form of EFT has a velocity/momentum regulator for light-cone divergencies, which can be used for any non-perturbative calculations. Through SCET, one can match it with other existing perturbative regulators.}
    
    \item{A Euclidean theory: the high-energy constants or momentum distributions of the above theory are not time-dependent. 
    Therefore, all calculations with Eq. (19) can be analytically continued to Euclidean time, and can be simulated, in particular, through lattice QCD. We now have a mature method to calculate these non-perturbative properties of hadrons.}
\end{itemize}
The above theory may be properly called large-velocity/rapidity EFT because formally the power counting is done in 
$1/\gamma$ and $1/{\bar v}\cdot D$. However, in practical calculations, it is the inverse hadron momentum $P^z$ appears in a power expansion ($M^2/(P^z)^2$ and $(1/xP^z)^2$), and for this reason, it has been named as large-momentum effective theory (LaMET)~\cite{Ji:2014gla}. 

\section{Expansion of Partons in Terms of Euclidean Distributions}

To summarize the above, the standard EFT of pQCD uses the light-cone formulation for partons, which has light-cone singularities and is Minkowskian in the sense that the time-dependence is essential,
and thus is difficult to implement for non-perturbative calculations. The alternative factorization leads to Euclidean correlators of collinear fields, and has a natural regulator amenable for non-perturbative calculations. Since these two EFTs are entirely equivalent for non-perturbative physics, it is natural to connect them to arrive at a LaMET expansion for parton physics~\cite{Ji:2013dva}. 

In fact, all light-cone distributions
properly defined in some perturbative UV and rapididty regularization scheme 
can be expanded in terms of generalized momentum distributions, in a way similar to PDFs, 
\begin{equation}
  f(x, \mu) = \int^\infty_{-\infty} \frac{dy}{y} C(y/x, P^z/\mu)\tilde f(y, P^z)  \ .  
\end{equation}
in the $P^z\to \infty$ limit. 
This is simple to see for perturbative quark/gluon states, in which both $f$ and $\tilde f$ can be calculated to all orders in perturbation theory and their infrared structures are exactly the same. The only difference is in the UV and can be matched up by the perturbative Wilson coefficient, $C$. Operationally, $C$ can be obtained from a forest formula similar to the BPHZ renormalization process~\cite{Itzykson:1980rh,Zimmermann:1969jj}. 

\subsection{LaMET for partons in the sense of Weinberg}

A salient feature of EFTs is to calculate physical observables in a systematic, controlled expansion with some well-defined constants, not just in the $P^z\to \infty$ limit, but also at finite fixed $P^z$.

Due to the non-locality of the LaMET,
power corrections are arranged by twists of local operators and usually involve more complicated 
multi-field equal-time correlation functions. Twist-3 corrections to twist-2 parton distributions come from 
the Wilson line self-energy and can be taken into account through leading renormalon resummation~\cite{Zhang:2023bxs}. Twist-4 corrections are
dynamical. A typical twist-4 distribution is~\cite{Chen:2016utp} 
\begin{eqnarray}
   && \tilde f(y_1,y_2,y_3, P^z) 
   = \int^\infty_{-\infty} dz_1 dz_2 dz_3 \\ 
  && \times  \langle P|\overline{\psi}(z_1) 
   \gamma^z W(z_1,z_2) \psi(z_2)\overline{\psi}(z_3)  \gamma^z W(z_3, 0)\psi(0)|P\rangle \nonumber 
\end{eqnarray}
which is an equal-time, gauge-invariant correlation function along the $z$-direction. 

In the Weinberg's sense of an EFT, the parton observables can be treated as ``physical quantities", and unknown constants are momentum distributions or similar equal-time
correlation functions, which can be computed through lattice QCD simulations. Thus, one arrives at a full expansion formula
for PDFs at fixed $P^z$, 
\begin{eqnarray}
 &&  f(x, \mu) = \int^\infty_{-\infty} \frac{dy}{y} C_2\left(\frac{y}{x}, \frac{P^z}{\mu}\right)\tilde f\left(y, \frac{P^z}{\mu}\right) \nonumber \\
  && + \left(\frac{\Lambda_{\rm QCD} }{P^z}\right)^2 \sum_i 
  \int^\infty_{-\infty} \frac{dy_1}{y_1}  \frac{dy_2}{y_2}  \frac{dy_3}{y_3} C_{4i}\left(\frac{y_1}{x}, \frac{y_2}{x},\frac{y_3}{x}, \frac{P^z}{\mu}\right) \nonumber \\ 
 && \times \tilde f_i\left(y_1,y_2,y_3,\frac{{P^z}}{\mu} \right) + ...
\end{eqnarray}
Technically, all power corrections have power divergences and the subtractions must be defined consistently with regularization of the IR renormalon poles in the twist-2 $C_2$~\cite{tHooft:1977xjm,Beneke:1998ui}. 

Therefore, the $x$-dependence of PDFs can be calculated with systematically-controlled precision from the above equation. There is no inverse problem as in parton phenomenology and short-distance expansion where the $x$-dependence is assumed to be a combination of some model functions.

\subsection{Analogy to Symanzik's improved/perfect action in lattice QCD}

Continuum field theories generally have UV singularities, and are not well-defined unless there is a proper cut-off. The well-known dimensional regularization works only for perturbative calculations in which Feynman integrals can be performed in $D=4-\epsilon$ dimension. 

Lattice QCD regulates the UV divergences by replacing
space-time with a lattice of spacing $a$.
All IR degrees of freedoms are kept but UV degrees of freedoms with momentum larger than $\pi/a$ are cut off and recovered in the continuum $a\to 0$ limit. Alternatively, one may consider the UV degrees of freedoms to have been integrated out in perturbation theory according to Wilson's renormalization group method. The result is a Symanzik's lattice effective theory with an 
lagrangian~\cite{Symanzik:1983dc,Symanzik:1983gh},  
\begin{equation}
  {\cal L}_{\rm eff}^{\rm QCD} = C_0(\alpha(a)){\cal L}^{(0)}_a + \sum_{i=1}^\infty a^i C_i(\alpha(a)){\cal L}^{(i)}_a \ , 
\end{equation}
where ${\cal L}^{(0)}$ is the naive discretized lagrangian density from the continuum one.  ${\cal L}^{(i)}$ are higher dimensional operators generating norminal power corrections. This ``perfect action'' will
eliminated any lattice-spacing dependence in a physical observable. 

Similarly, the $v\to c$ limit of parton theory is not well defined due to light-cone divergences.  But, one can work at a fixed large $v$ as a cut-off. Physics missing between the cut-off and light-cone can be regained through a full LaMET expansion of the effective lagrangian, 
\begin{equation}
  {\cal L}_{\rm eff}^{\rm Parton} = C_0(\alpha){\cal L}^{(0)}_v + \sum_{i=1}^\infty \gamma^{-i} C_i(\alpha){\cal L}^{(i)}_v \ .     
\end{equation}
where we have used only $1/\gamma$ to indicate power corrections and non-local terms with $1/in\cdot D$ have been omitted. For computing light-cone PDFs, $P =M\gamma v$ shows up when a particular hadron state of mass $M$ and velocity $v$ is involved. While Wilson coefficient $C_2$ helps to eliminate all the logarithmic dependences in $P$, $P$-dependence is completely removed through the infinite power corrections, and is finally replaced by the UV 
and rapidity renormalization scales $\mu$ and $\zeta$ in light-cone matrix elements which can be defined through perturbative schemes.

It shall be warned that the effective parton theory is only effective when all parton momenta are at perturbative scales, $xP\gg\Lambda_{\rm QCD}$. For example, the above expansion breaks down at small $x\sim \Lambda_{\rm QCD}/P$.  

\subsection{PDF and GPD singularities}

Light-cone PDFs such as $f(x)$ are singular functions of parton momentum fraction $x$. They vanish at $x=1$ but their first and high-order derivatives are not continuous. Moreover, at $x=0$ PDFs are divergent. 

In LaMET expansion for PDFs, momentum distribution $\tilde f(y, P^z)$ is analytic in whole range of $y$ at finite $P^z$ including $y=1$. This is because with backward moving particles in a finite momentum hadron, a quark or gluon can carry more momentum than the total $P^z$. Since the correlation function in coordinate space $z$ has finite support (decays exponentially at
large $z$) due to confinement, the momentum distribution is analytical at both $y=0$ or $y=1$. At finite $P^z$, LaMET expansion for $f(x)$ breaks down near $x=0$ and $1$, and therefore cannot give any useful information about PDF singularities. 

On the other hand, in the limit of $P^z$ going to infinity, $C_2$ increasingly approaches $\delta(x/y-1)$,  and the momentum distribution $\tilde f(y, P^z)$ approaches PDF $f(x, \mu)$ with a proper scheme conversion.  The correlation length $\lambda = zP^z$ diverges, therefore $\tilde f(y,\infty)$ becomes singular at $y=x=0$. Moreover, there is no more backward moving quarks and gluons in the hadron in this limit, and  $\tilde f(1, \infty)$ no longer has any non-vanishing support. But the derivatives of $\tilde f(y, \infty)$ at $y=1$ need not vanish. 

Besides the singularities at $x=\pm 1$, generalized parton distributions (GPDs) have singularities at $x=\pm \xi$ ( $\xi$ is a skewness variable) where one of the parton has zero momentum fraction~\cite{Muller:1994ses,Ji:1996ek}. GPDs are continuous there but derivatives are not. 
One can generate these singularities through the same 
mechanism discussed above.  

In a recent paper~\cite{Braun:2024snf}, it has been suggested that due to the renormalon effect, a linear power correction in factorization of quasi GPDs maybe generated from convoluting the coefficient functions and light-cone GPDs because of the derivative discontinuity of the latter at $x=\pm \xi$. If, on the other hand, the light-cone GPDs are factorized in terms of the quasi GPDs, the coefficient functions will be convoluted with smooth quasi-GPDs, and nothing special seems to happen at  $x=\pm \xi$. It remains to see if a special linear power correction indeed disappears at this point. 
\\

\section{Conclusion}

Weinberg's pioneering work on EFTs has had a great impact in modern theoretical physics. It changed fundamentally the way we think about quantum field theories. LaMET is an example of EFTs for calculating non-perturbative parton physics systematically through lattice QCD simulations~\cite{Ji:2020ect}, 
which was thought impossible for a long while.   

\section*{Acknowledgments}

This work is partially supported by Maryland Center for Fundamental Physics. I thank Yushan Su, Feng Yuan and Yong Zhao for discussions related to the content of this paper, and Yushan Su for drawing the figure 1. I also thank Andres Schaefer for a careful reading of the manuscript and many useful suggestions for improvement. 

\bibliography{references}

\end{document}